\documentclass[11pt]{article}
\usepackage{fullpage}
\usepackage{lmodern}
\usepackage[english]{babel}
\usepackage[numbers]{natbib}
\usepackage{amsmath}
\usepackage{graphicx}
\usepackage{color}
\usepackage{hyperref}
\usepackage{authblk}
\usepackage{subcaption}
\renewcommand{\thefootnote}{\fnsymbol{footnote}}
\title{Constraints on electromagnetic properties of sterile neutrinos from MiniBooNE results}
\author[1,2]{A.~A.~Radionov\footnote{radionov@ms2.inr.ac.ru}}
\affil[1]{Institute for Nuclear Research of the Russian Academy of Sciences,\newline 60th October Anniversary Prospect, 7a, 117312 Moscow, Russia}
\affil[2]{Faculty of Physics, Moscow State University, Vorobjevy Gory, 119991 Moscow, Russia}
\begin{document}
\maketitle
\begin{abstract}
Among the class of models with small mixing angles between sterile and active neutrinos, we place constraints on the effective muon-to-sterile neutrino magnetic and electric dipole transition moments from the combined MiniBooNE results for the sterile neutrino mass range of $10\;\mathrm{MeV}<m_s<500\; \mathrm{MeV}$. Our results are valid for models with CP-violating interactions and for Dirac and Majorana sterile neutrinos. In addition, we show that such dipole electromagnetic interactions cannot be the main source of the anomalous events in the MiniBooNE experiment because they fail to reproduce the anomalous event distribution as a function of polar angle. However, good agreement with the anomalous event distribution in reconstructed energy can be achieved for some values of magnetic and electric moments.
\end{abstract}
\renewcommand{\thefootnote}{\arabic{footnote}}
\section{Introduction}\label{sectionintro}
In recent decades, significant progress has been achieved in experimental studies of neutrinos. Because of the incompleteness of  the Standard Model of particle physics particularly in this field, many questions about neutrino properties have arisen. The Mini--Booster Neutrino Experiment (MiniBooNE) has observed a so-called "anomaly": a statistically significant excess of detected events at low energies in comparison with theoretical predictions \cite{miniboone0812excess,miniboone1007excess,miniboone1207excess}. Many efforts have already been made and several hypotheses have been put forward to explain this phenomenon (see the list in, e.g., \cite{miniboone1207excess}).

One of the hypotheses consists of introducing a sterile neutrino with mass in the range 40-600 MeV that is unstable with respect to radiative decay. There are two realizations of this idea in the literature. First, sterile neutrinos could be produced by flavor mixing with active (muon) neutrinos \cite{gninenko,gninenko12} by scattering from nuclei due to neutral--current weak interactions ($\nu_{\mu}\rightarrow\nu_s$). This process is followed by subsequent radiative decay due to a transition moment ($\nu_s\rightarrow\nu\gamma$). This idea has already been tested experimentally with negative results \cite{duk2012}. Thus, only the mass region from 400-600 MeV \cite{gninenko2009} remains allowable, and this hypothesis waits for special analyses of data from c- and b-factories \cite{gninenkogorbunov0907}. As for the second explanation, it was argued \cite{masip1210} that the dipole transition moment may be responsible for both production and decay of a sterile neutrino. In this case, it would be a dipole transition moment between sterile and muon neutrinos that would do the main job. Both of these suggested explanations have renewed interest in dipole transition moments of hypothetical heavy sterile neutrinos.

In this paper, we extract new and improve old  \cite{gninenkokrasnikov1998} limits on the values of sterile-to-muon neutrino transition dipole moments under conditions in which we neglect flavor mixing between sterile and active neutrinos. To this end, we analyze the results of the MiniBooNE detector \cite{miniboone0806detector}, which measures Charge-Current-like and Neutral-Current-like events from neutrino and antineutrino fluxes. These fluxes are mainly composed of muon neutrinos and antineutrinos with energies in the range $200\;\mathrm{MeV}<E_{\nu}<3\;\mathrm{GeV}$. The detector cannot distinguish photon events from electron/positron events \cite{miniboone0812excess}, so neutrino transition dipole moments can be probed because transition moments lead to photoproduction in the detector. 

The paper is organized as follows. Section \ref{sectionmodel} describes the basics of neutrino transition dipole moments and introduces useful parameterizations of  them; then, we calculate a cross-section of active-to-sterile (anti)neutrino conversion on an atom and  sterile (anti)neutrino radiative decay rates. Section \ref{sectionpred} contains a probability function of sterile neutrino decay inside the MiniBooNE detector and the distributions of the expected photons in energy and polar angle. In particular, we show here that neutrinos produced on nuclei and atomic electrons via the exchange of massless particles (photons) are mostly forward-directed, as are the photons from their decay. These results fail to reproduce the observed excess as a function of polar angle \cite{miniboone0812excess,miniboone1007excess}. Section \ref{sectionanalysisandresults} describes the methods we exploit to put constraints on dipole transition moments, summarizes our results and describes extensions of the model that may (possibly) be relevant for MiniBooNE.

\section{Model description}\label{sectionmodel}

Here, we describe a model using Dirac sterile neutrinos. The case of Majorana neutrinos requires minor modifications, which are listed at the end of this Section. 
	
The  most general Lorenz-invariant electromagnetic dipole interaction between neutrinos and an electromagnetic field is
\begin{equation}\label{inter0}
\mathcal{L}_{int}=\frac{1}{2}\sum_{\substack{i,j=1\\i\leq j}}^{N_{\nu}}\overline{\nu}_j\left(\mu_{ij}+\gamma_5d_{ij}\right)\sigma_{\mu\nu}\mathcal{F}^{\mu\nu}\nu_{i} + \mathrm{h.c.}\;,
\end{equation}
where $\nu_i$ are neutrino fields in the mass basis, $N_{\nu}$ is a number of neutrinos, $\mu^{tr}_{ij}\equiv \mu_{ij}\left(d^{tr}_{ij} \equiv d_{ij}\right)$ with $i\neq j$ are magnetic (electric) transition dipole  moments between massive neutrinos $i$ and $j$, $\mu_{ii}$ ($d_{ii}$) are diagonal magnetic (electric) transition dipole moments,  $\mathcal{F}^{\mu\nu}=\partial^{\mu}\!A^{\nu}-\partial^{\nu}\!A^{\mu}$ is an electromagnetic field strength tensor and $\sigma_{\mu\nu}=\frac{i}{2}\left(\gamma_\mu\gamma_\nu-\gamma_\nu\gamma_\mu\right)$ where $\gamma_\mu$ are the Dirac matrices. We investigate a model with $N_{\nu}=4$, so we suppose the existence of one additional sterile neutrino\footnote{We extend the analysis to the case of $N_{\nu}>4$ case in Section \ref{sectionanalysisandresults}.} with mass $m_s$. 

For our study, it is convenient to work with gauge (flavor) states in the {\em active neutrino sector}. A sterile neutrino $\nu_s$ is a massive fermion that is neutral with respect to the Standard Model gauge group. We are interested in a situation where flavor mixing between sterile and active neutrinos is negligible so that the sterile neutrino in flavor basis $\nu_s$ and the heaviest neutrino in mass basis $\nu_4$ are almost the same: $\nu_4\simeq \nu_s$. We denote the mass of a sterile neutrino by $m_s$.

\begin{figure}[h!]
\centering\fbox{\includegraphics[scale=2.0]{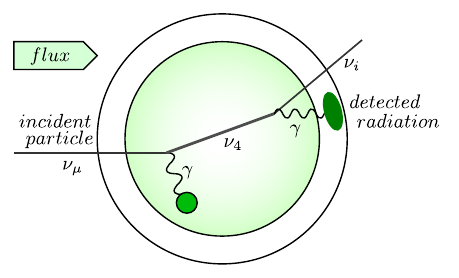}}
\caption{{\small Schematic illustration of the conversion and decay of $\nu_4$ in the MiniBooNE detector.}}
\label{figsphere}
\end{figure}

Interaction \eqref{inter0} would induce two processes: an active-sterile neutrino conversion on atomic electrons and nuclei and sterile neutrino decay back to an active neutrino (plus a photon) [see Fig.\ref{figsphere}]. 

The transition dipole moments entering \eqref{inter0} are related to dipole moments in the flavor basis through the following relations:
\begin{equation}
\begin{array}{c}
\mu_{\alpha\beta}=\sum_{i,j=1}^4{\mu_{ij}U^*_{\alpha i}U_{\beta j}}\;,\\
d_{\alpha\beta}=\sum_{i,j=1}^4{d_{ij}U^*_{\alpha i}U_{\beta j}}\;,
\end{array}
\end{equation} 
where $U_{\alpha i}$ is a generalization of the Pontecorvo--Maki--Nakagawa--Sakata matrix,
\begin{equation}
\nu_{\alpha}\left(x\right)=\sum_{i}U_{\alpha i}\nu_{i}\left(x\right)\; .
\end{equation} 
It is convenient also to define these quantities in another basis:
\begin{equation}
\begin{array}{c}
\mu_{\alpha j}=\sum_{i=1}^4{\mu_{ij}U^*_{\alpha i}}\;,\\
d_{\alpha j}=\sum_{i=1}^4{d_{ij}U^*_{\alpha i}}\;.
\end{array}
\end{equation} 
In this paper, we are mostly interested in the following term:
\begin{equation}\label{inter1}
\mathcal{L}^{\prime}_{int}=\frac{1}{2}\overline{\nu}_4\left(\mu^{tr}_{\mu 4}+\gamma_5d^{tr}_{\mu 4}\right)\sigma_{\rho\lambda}\mathcal{F}^{\rho\lambda}\nu_{\mu} + \mathrm{h.c.}\;.
\end{equation}
The analysis described below allows us to place constraints on the quantity
\begin{equation}\label{kappadef}
\kappa^{tr}_{\mu 4}\equiv\sqrt{\vert \mu^{tr}_{\mu 4}\vert^2+\vert d^{tr}_{\mu 4}\vert^2}\;.
\end{equation}
Generally, electric and magnetic moments are complex quantities that have different phases. We parameterize them in the form
\begin{equation}\label{lpcdef}
 \mu^{tr}_{\mu 4}\equiv\kappa^{tr}_{\mu 4}\cos{\left(\lambda^{tr}_{\mu 4}\right)}\exp{\left(i\phi^{tr}_{\mu 4}\right)}~~~ \mathrm{and}~~~  d^{tr}_{\mu 4}\equiv\kappa^{tr}_{\mu 4}\sin{\left(\lambda^{tr}_{\mu 4}\right)}\exp{\left(i\chi^{tr}_{\mu 4}\right)}\;,
\end{equation}
where $\lambda^{tr}_{\mu 4}$,  $\phi^{tr}_{\mu 4}$ and $\chi^{tr}_{\mu 4}$ are real-valued parameters. Analogous notations are adopted for transition moments in the mass basis:
\begin{equation}
 \mu^{tr}_{4i}\equiv\kappa^{tr}_{4i}\cos{\left(\lambda^{tr}_{4i}\right)}\exp{\left(i\phi^{tr}_{4i}\right)}~~~ \mathrm{and}~~~ d^{tr}_{4i}\equiv\kappa^{tr}_{4i}\sin{\left(\lambda^{tr}_{4i}\right)}\exp{\left(i\chi^{tr}_{4i}\right)}\;,
\end{equation}
where $\lambda^{tr}_{4i}$,  $\phi^{tr}_{4i}$ and $\chi^{tr}_{4i}$ are real-valued parameters.

In the MiniBooNE detector, the conversion of muon (anti)neutrinos to sterile (anti)neutrinos would happen mostly on carbon atoms bound up in oil. In the quasi-elastic approximation, the conversion cross-section contains a form factor that depends on a single parameter $t=-q^2$ (where $q$ is a 4-momentum transfer carried by the photon) as given in \cite{tsai1974}. The form factor  \cite{tsai1974} was used in previous studies of sterile neutrino magnetic transition  moments \cite{masip1210,gninenkokrasnikov1998}. (In Ref.\;\cite{masip1210}, a different nuclear part of the form factor was adopted for large $t>10^{-3}\;\mathrm{GeV^2}$; however, the main contribution comes from atomic and coherent nuclear processes where $t$ is smaller, so the difference is insignificant.) In the case of a light nucleus (carbon: atomic number and atomic mass of $Z=6$ and $A=12$, respectively), an analytical expression for the form factor is \cite{tsai1974}
\begin{equation}\label{formfact}
G^2\left(t\right)=
\begin{cases}
\frac{Z^2a^4t^2}{\left(1+a^2t\right)^2}+\frac{Za^{\prime 4}t^2}{\left(1+a^{\prime 2}t\right)^2}, &  t<t_{0};\;\text{(atomic part of the form factor)}\;,\\ 
\frac{Z^2}{\left(1+\frac{t}{d}\right)^2}, &  t>t_{0};\;\text{(nuclear part of the form factor)}\;,
\end{cases}
\end{equation}
where $a=184.15\times\left(2.718\right)^{-1/2}Z^{-1/3}/m_e$,  $a^{\prime}=1194\times\left(2.718\right)^{-1/2}Z^{-2/3}/m_e$ and $d=0.164\times A^{-2/3} \mathrm{GeV}^2$ \cite{tsai1974}; $t_0=7.39\times m_e^2$ \cite{gninenkokrasnikov1998} ($m_e$ is electron mass).

For the differential cross-sections of muon neutrino conversion into sterile neutrinos of positive $d\sigma_{+,\nu}$ and negative $d\sigma_{-,\nu}$ chirality, we obtain
\begin{equation}\label{convproc1}
\frac{d\sigma^{\nu}_{\mp}\left(E,\theta,\phi\right)}{d\cos\theta d\phi}=\frac{
\alpha\left(\kappa^{tr}_{\mu 4}\right)^2}{4\pi}\left(1-\sin\left(2 \lambda^{tr}_{\mu 4}\right)\cos{\left( \phi^{tr}_{\mu 4}- \chi^{tr}_{\mu 4}\right)}\right)\frac{G^2\left(t\right)}{t^2}vE^4\left(1\pm v\right)^3\left(1\mp\cos\theta\right)\;,\\
\end{equation}
where $\kappa^{tr}_{\mu 4}$, $\lambda^{tr}_{\mu 4}$, $\phi^{tr}_{\mu 4}$ and $\chi^{tr}_{\mu 4}$ are defined in Eqs.\;\eqref{kappadef} and \eqref{lpcdef}, $E$ is the incident muon neutrino energy, $v=\sqrt{E^2-m_s^2}/E$ and $\theta$ is the angle between the muon neutrino 3-momentum and the sterile neutrino 3-momentum in the laboratory frame. Equation \eqref{convproc1} generalizes Eq.\;(4) of Ref.\;\cite{gninenkokrasnikov1998} to the case of models where electric and magnetic transition moments are presented.

For muon antineutrinos, the conversion cross-section is
\begin{equation}\label{convproc2}
\frac{d\sigma^{\bar{\nu}}_{\mp}\left(E,\theta,\phi\right)}{d\cos\theta d\phi}=\frac{
\alpha\left(\kappa^{tr}_{\mu 4}\right)^2}{4\pi}\left(1+\sin\left(2 \lambda^{tr}_{\mu 4}\right)\cos{\left( \phi^{tr}_{\mu 4}- \chi^{tr}_{\mu 4}\right)}\right)\frac{G^2\left(t\right)}{t^2}vE^4\left(1\pm v\right)^3\left(1\mp\cos\theta\right)\;.
\end{equation}

We see that neutrino and antineutrino cross-sections complement each other in the following sense: at least one of them is non-zero for any choice of phases $\lambda^{tr}_{\mu 4}$, $\phi^{tr}_{\mu 4}$ and $\chi^{tr}_{\mu 4}$ if $\kappa^{tr}_{\mu 4}$ is non-zero. Combined with the analogous property of sterile neutrino decay rate (see Eqs.\;\eqref{decayproba} and \eqref{decayprobb}), it would follow that we can limit the value of $\kappa_{\mu 4}^{tr}$. Note in passing that in Eqs.\;\eqref{convproc1} and \eqref{convproc2} we assume that quasi-elastic processes dominate and the energy of the heavy neutrino is equal to the incident muon neutrino energy.

Let us proceed with a description of heavy sterile (anti)neutrino radiative decay. The (anti)neutrino decays into some other (anti)neutrino of significantly smaller\footnote{It is important that final (anti)neutrinos have significantly smaller mass in comparison with $m_s$ or the spectra get shifted towards smaller energies and all further calculations have to be modified as in Section \ref{sectionanalysisandresults}.} mass $\left( m_j \ll m_s  \right)$ and a photon. Formulas for differential decay rates of sterile (anti)neutrinos of a given chirality ($\pm$) are
\begin{equation}\label{decayproba}
\frac{d\Gamma^{\nu}_{\pm}}{d\phi_{\gamma}d\cos\theta_{\gamma}}=\frac{1}{32\pi^2}m_s^3\sum_{i=1}^3\left(\kappa^{tr}_{4i}\right)^2\left(1\mp\sin\left(2 \lambda^{tr}_{4i}\right)\cos{\left( \phi^{tr}_{4i}- \chi^{tr}_{4i}\right)}\cos{\left(\theta_\gamma\right)}\right)\;,
\end{equation}
\begin{equation}\label{decayprobb}
\frac{d\Gamma^{\overline{\nu}}_{\pm}}{d\phi_{\gamma}d\cos\theta_{\gamma}}=\frac{1}{32\pi^2}m_s^3\sum_{i=1}^3\left(\kappa^{tr}_{4i}\right)^2\left(1\pm\sin\left(2 \lambda^{tr}_{4i}\right)\cos{\left( \phi^{tr}_{4i}- \chi^{tr}_{4i}\right)}\cos{\left(\theta_\gamma\right)}\right)\;,
\end{equation}
where $\theta_{\gamma}$ and $\phi_{\gamma}$ are polar and azimuthal angles of the out-coming photon's 3-momentum in the sterile neutrino rest frame measured from the 3-momentum of the heavy neutrino in the laboratory frame. As we see, generally there is an anisotropy in polar angle in Eqs.\;\eqref{decayproba} and \eqref{decayprobb}.

The full width is (see, e.g., \cite{mohapatra})
\begin{equation}
\Gamma=\Gamma^{\nu}_{\pm}=\Gamma^{\overline{\nu}}_{\pm}=\frac{1}{8\pi}m_s^3\sum_{i=1}^3\left(\kappa^{tr}_{4i}\right)^2\;.
\end{equation}

The difference between Majorana and Dirac neutrinos slightly modifies the formulas, and to get constraints for the Majorana case, one should substitute everywhere for $\kappa_{\mu 4}$ with $2\kappa_{\mu 4}$ and for $\kappa_{4i}$ with $2\kappa_{4i}$. In this paper, we present constraints for Dirac neutrinos. In the case of Majorana neutrinos, one should divide the obtained constraints on $\kappa_{\mu 4}$ for the Dirac case by a factor of two.

\section{Theoretical predictions}\label{sectionpred}

In this Section, we derive approximate analytical formulas for the number of photon events associated with sterile neutrinos that should be observed in the MiniBooNE detector.  
The MiniBooNE experiment has two operating modes: $\boldsymbol{mode\in\{neutrino,antineutrino\}}$. In the neutrino mode, there is a dominant flux of muon neutrinos and a subdominant flux of muon antineutrinos; in the antineutrino mode, there is a dominant flux of muon antineutrinos and a subdominant flux of muon neutrinos \cite{miniboone0806flux}. (Electron neutrinos do not play a role in our analysis, and  they are neglected in what follows.) It should be emphasized that these two fluxes would be independent sources of photons. 

We use the following notations for angle coordinates: ($\theta$,$\phi$) are for converted heavy neutrinos in the laboratory frame, ($\theta_{\gamma}$,$\phi_{\gamma}$) are for photons in the heavy neutrino rest frame and  ($\theta_{det}$,$\phi_{det}$) are for photons in the laboratory frame.

Bins to measure distributions of reconstructed quasi-elastic energy events (evaluated by observed energy and polar angle, see Eq.\;(3) of Ref.\;\cite{miniboone0706}) are
 \cite{miniboone0812excess,miniboone1007excess,miniboone1207excess}: 
\begin{equation}
\boldsymbol{bin=[E^{QE}_{min},E^{QE}_{max}]\in\{[0.2\;\mathrm{GeV},0.3\;\mathrm{GeV}],\;[0.3\;\mathrm{GeV},0.375\;\mathrm{GeV}],\; \mathrm{etc.}\}}\;.
\end{equation}
Observed energy bins are \cite{miniboone1007excess,miniboone1207excess}
\begin{equation}
\boldsymbol{bin=[E_{min},E_{max}]\in\{[0.1\;\mathrm{GeV},0.2\;\mathrm{GeV}],\;[0.2\;\mathrm{GeV}, 0.3\;\mathrm{GeV}],\; \mathrm{etc.}\}}\;.
\end{equation}
Polar angle bins are \cite{miniboone0812excess,miniboone1007excess}
\begin{equation}
\boldsymbol{bin=[\cos{\theta_{min}},\cos{\theta_{max}}]\in\{[-1.0,-0.8],\;[-0.8,-0.6],\; \mathrm{etc.}\}}\;.
\end{equation}

\begin{figure}[h!]
\centering\includegraphics{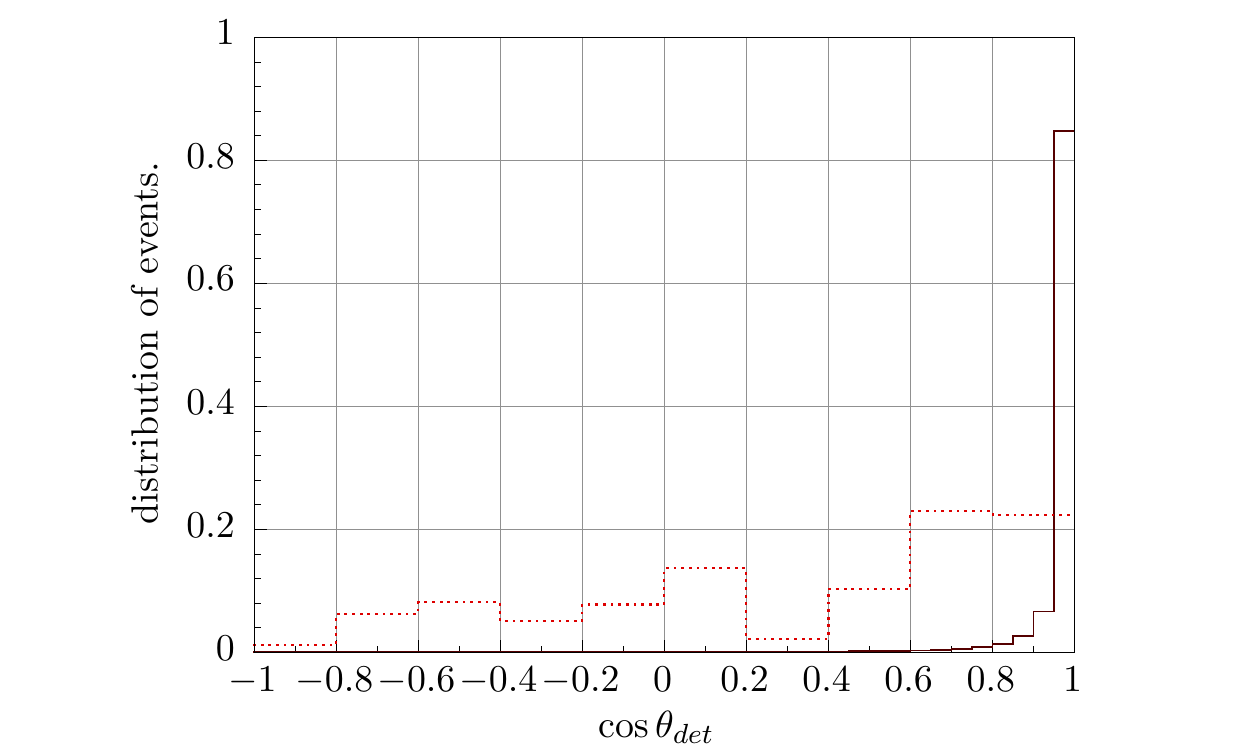}
\caption{{\small Theoretically predicted events as a function of $\cos\theta_{det}$ in the neutrino operating mode for the parameters \newline $m_s=50\;\mathrm{MeV}$, $\kappa_{\mu 4}^{tr}=1\times10^{-9}\mu_B$, $\tau=1/\Gamma=1.5\times10^{-6}\;\mathrm{s}$ , $\sin\left(2\lambda_{\mu}\right)=0$ and $\sin{\left(2\lambda_4\right)}=0$ (solid line) and the observed excess (the dashed line) \cite{miniboone1207excess}.}}
\label{figangulardist}
\end{figure}

In each bin, we compare the number of events expected due to dipole interactions with the number of {\it anomalous events} (the excess is defined as the difference between numbers of events observed and events predicted by the Standard Model with three active massive neutrinos). For the energy range of detected photons under investigation, the predictions depend monotonically on the energy of detected photons. The differences between the observed and reconstructed quasi-elastic energies of the photons  are negative and negligible for photons at small polar angles $\theta_{det}$ in comparison with the detector resolution and bin sizes. Therefore, it is reasonable to neglect the difference between observed and reconstructed energies because almost all predicted events have very small polar angle $\theta_{det}$ (see Fig.\;\ref{figangulardist}). 

The efficiency of the MiniBooNE detector $\epsilon_{\gamma}\left(E\right)$ at registering photons is determined as a piecewise function of observed energy \cite{miniboone1207excess}. For the energy distributions, we set the efficiency inside each bin as corresponding to a constant. For polar angle distribution, we set the efficiency to its minimal value $\epsilon_{\gamma}=7.3\%$ to simplify the calculations when we work on constraints (this gives conservative constraints for photon energies below $E_{\gamma}<1.5\;\mathrm{GeV}$). These measures are justified by a very moderate dependence of the results (i.e., the limits on $\kappa^{tr}_{\mu 4}$) on the variation in efficiency.

To be observed, sterile neutrinos  must decay inside a sphere of radius $R=5.0\;\mathrm{m}$. The MiniBooNE detector is located far away from the beam target, so heavy (anti)neutrino fluxes are approximately homogeneous over the volume of the spherical detector. For sterile neutrinos of energy $E$, we define the average probability of a decay inside the inner sphere. Because heavy neutrinos could be produced anywhere inside the inner sphere, this probability is evaluated by averaging the probability for a heavy neutrino to decay inside the total volume:
\begin{equation}\label{decint}
P\left(x\right)=\frac{1}{4/3\pi R^3}\int_0^R 2\pi r dr \int_0^{2\sqrt{R^2-r^2}}\left(1-\exp{\left(\frac{-xz}{R}\right)}\right)dz\;,
\end{equation}
where a dimensionless parameter $x$  is defined as the product of the inverse decay length and the radius of the inner sphere as
\begin{equation}
x=\frac{m_s^3\kappa^2}{8\pi}\frac{m_s}{\sqrt{E^2-m_s^2}}R\;.
\end{equation}
Integral \eqref{decint} is equal to
\begin{equation}\label{averdecay}
P\left(x\right)=1 - \frac{3\left(2 x^2 - 1 + \left(2x + 1\right)e^{-2x}\right)}{8x^3}\;.
\end{equation}

Below, we obtain theoretical predictions for the distributions of observed energy and polar angle, and we discuss the approximations we used in calculating the constraints.

From Eqs.\;\eqref{averdecay}, \eqref{convproc1}, \eqref{decayproba}, \eqref{convproc2} and \eqref{decayprobb}, we give an analytical estimate for the number of events in both operating modes of the MiniBooNE experiment. We introduce two functions: the first corresponds to the isotropic part of the differential decay rates \eqref{decayproba} and \eqref{decayprobb} (the $\theta_{\gamma}$-independent terms)
\begin{equation}
\begin{split}
\mathcal{A}^{mode,particle}_{[E_{min},E_{max}]}=\left(\kappa_{\mu 4}^{tr}\right)^2\frac{\alpha}{4}\mathcal{N}^{mode}_{p.o.t.}N_{C}\epsilon_{\gamma}\int\!dE\,d\!\cos{\theta}E^3\sqrt{E^2-m_s^2}\rho_{mode}^{particle}\!\left(E\right)\frac{G^2\left(t\right)}{t^2}P\left(x\right)\times \\ \times \left(\left(1+ v\right)^3\left(1-\cos\theta\right)+\left(1-v\right)^3\left(1+\cos\theta\right)\right)\left(\cos\theta_{cut}\!\left(E_{max}\right)-\cos\theta_{cut}\!\left(E_{min}\right)\right)\;;
\end{split}
\end{equation}
while the second corresponds to the anisotropic part of the differential decay rates \eqref{decayproba} and \eqref{decayprobb} (the $\theta_{\gamma}$-dependent terms):
\begin{equation}
\begin{split}
\mathcal{B}^{mode,particle}_{[E_{min},E_{max}]}=\left(\kappa_{\mu 4}^{tr}\right)^2 \frac{\alpha}{8}\mathcal{N}^{mode}_{p.o.t.}N_{C}\epsilon_{\gamma}\int\!dE\,d\!\cos{\theta}E^3\sqrt{E^2-m_s^2}\rho_{mode}^{particle}\!\left(E\right)\frac{G^2\left(t\right)}{t^2}P\left(x\right)\times \\ \times \left(\left(1+ v\right)^3\left(1-\cos\theta\right)-\left(1-v\right)^3\left(1+\cos\theta\right)\right)\left(\cos^2\theta_{cut}\!\left(E_{max}\right)-\cos^2\theta_{cut}\!\left(E_{min}\right)\right)\;,
\end{split}
\end{equation}
where $\rho^{\left(anti\right)neutrino}_{mode}\!\left(E\right)$ is a muon (anti)neutrino energy spectrum in a given operating mode \cite{miniboone0806flux}, $\epsilon_{\gamma}=\epsilon_{\gamma}\!\left(0.5\left(E_{min}+E_{max}\right)\right)$ is the detection efficiency, $\mathcal{N}^{mode}_{p.o.t.}$ is the number of protons on target,  $N_C=\left(5.0/6.1\right)^3\times3.5\times10^{31}$ is the number of carbon atoms inside the inner sphere and the function $\cos\theta_{cut}\!\left(E\right)$ is defined as:
\begin{equation}\label{cutangle1}
\cos\theta_{cut}\!\left(E\right)=
\begin{cases}
-1, &\mbox{at}\hskip0.5cm E<\frac{m_s}{2}\sqrt{\frac{1-v}{1+v}} \\
\frac{1}{v}\left(\frac{2E}{m_s}\sqrt{1-v^2}-1\right), &\mbox{at }\hskip0.8cm\hskip-0.5cm \frac{m_s}{2}\sqrt{\frac{1-v}{1+v}}\leq E\leq\frac{m_s}{2}\sqrt{\frac{1+v}{1-v}} \\
1, &\mbox{at}\hskip0.5cm E>\frac{m_s}{2}\sqrt{\frac{1+v}{1-v}}\;.
\end{cases}
\end{equation}

In a given bin, the predicted number of events is defined as:
\begin{equation}\label{numberofevents}
\begin{split}
\mathcal{N}^{mode}_{[E_{min},E_{max}]}=\left(1-\sin{\left(2\lambda_{\mu}\right)}\right)\left(\mathcal{A}^{mode, neutrino}_{[E_{min},E_{max}]}-\sin{\left(2\lambda_{4}\right)}\mathcal{B}^{mode,neutrino}_{[E_{min},E_{max}]}\right)+\\
+\left(1+\sin{\left(2\lambda_{\mu}\right)}\right)\left(\mathcal{A}^{mode, antineutrino}_{[E_{min},E_{max}]}+\sin{\left(2\lambda_{4}\right)}\mathcal{B}^{mode,antineutrino}_{[E_{min},E_{max}]}\right)\;,
\end{split}
\end{equation}
where we introduced the variables
\begin{equation}
\sin{\left(2\lambda_{\mu}\right)}\equiv\sin{\left(2\lambda^{tr}_{\mu 4}\right)}\cos{\left( \phi^{tr}_{\mu 4}- \chi^{tr}_{\mu 4}\right)}~~\mathrm{and}~~\sin\left(2\lambda_4\right)\equiv\sum_{i=1}^3\frac{\left(\kappa^{tr}_{4i}\right)^2}{\kappa^2}\sin\left(2\lambda_{4i}\right)\cos{\left( \phi^{tr}_{4i}- \chi^{tr}_{4i}\right)}\;.
\end{equation}
Note that if $\cos{\theta_{cut}\!\left(E_{min}\right)}<\cos{\theta_{cut}\!\left(E_{max}\right)}$ then
\begin{equation}
\vert \mathcal{B}^{mode,particle}_{[E_{min},E_{max}]}\vert <  \mathcal{A}^{mode,particle}_{[E_{min},E_{max}]}\;,
\end{equation}
and, consequently, the predicted number of events (see Eq.\;\eqref{numberofevents}) would be non-zero for all values of parameters $\lambda_{\mu}$ and $\lambda_{4}$.

It is important that the theoretically predicted number of events \eqref{numberofevents} depends{\em monotonically}on parameters $\kappa_{\mu 4}^{tr}$ and $\kappa_{i4}^{tr}$. Provided the approximate relation
\begin{equation}\label{approxrel}
\sum_{i=1}^3\left(\kappa^{tr}_{4i}\right)^2 \geq \left(\kappa^{tr}_{\mu 4} \right)^2\;,
\end{equation}
which is valid up to the neglected flavor mixing between active and sterile neutrinos, we place constraints from above on the possible values of $\kappa^{tr}_{\mu 4}$. Indeed, if we substitute everywhere the combination $\sum_{i=1}^3\left(\kappa_{4i}^{tr}\right)^2$ by $\left(\kappa_{\mu 4}^{tr}\right)^2$, it would only suppress the predicted number of events \eqref{numberofevents}. Otherwise, if relation \eqref{approxrel} is invalid and, hence, flavor mixing between sterile and active neutrinos is significant, the constraints will depend not only on the mass of the sterile neutrino but on its lifetime as well.

\begin{figure}[!h]
\begin{tabular}{cc}
 \includegraphics{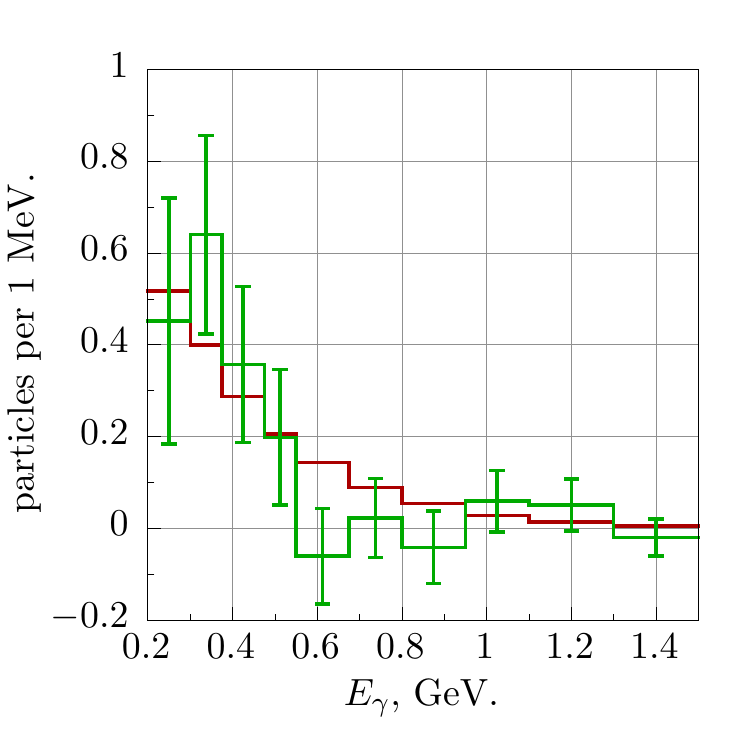} & \includegraphics{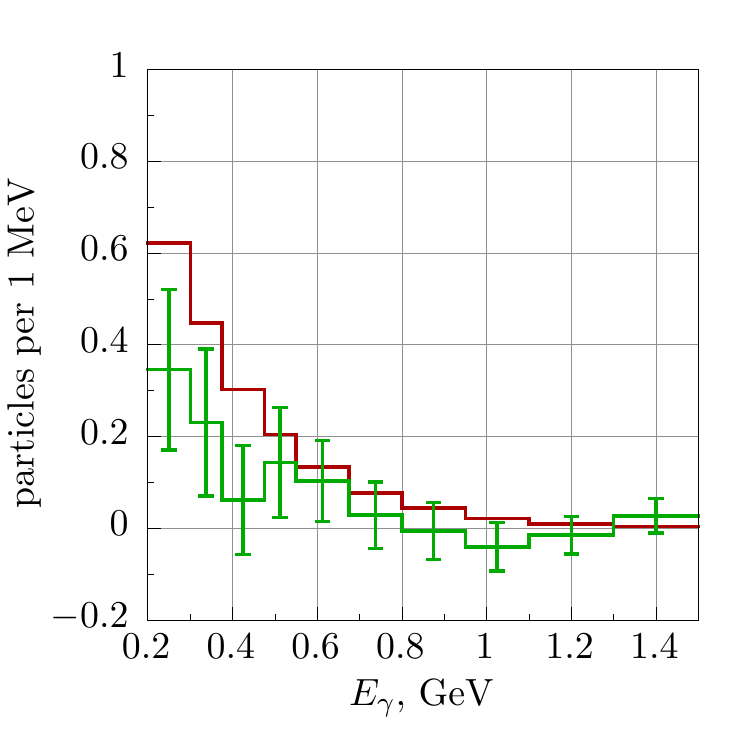}
\end{tabular}
\caption{{\small Energy distributions of excess events obtained in neutrino (left) and antineutrino (right) operating modes with error bars \cite{miniboone1207excess} and the corresponding predicted spectra of photons $\mathcal{N}^{mode}_{[E_{min},E_{max}]}$ from sterile neutrino decays for the parameter values $m_s=50\;\mathrm{MeV}$, $\kappa^{tr}_{\mu 4}=9.9\times 10^{-9} \mu_B$, $\tau=1/\Gamma=1.5\times10^{-8}\;s$, $\sin{\left(2\lambda_4\right)}=0$ and $\sin{\left(2\lambda_{\mu}\right)}=0$.}}
\label{enprofile}
\end{figure}

In Fig.\;\ref{enprofile}, we compare the predicted number of events $\mathcal{N}^{mode}_{[E_{min},E_{max}]}$ at  parameter values $m_s=50\;\mathrm{MeV}$, $\kappa^{tr}_{\mu 4}=9.9\times 10^{-9} \mu_B$, $\tau=1/\Gamma=1.5\times10^{-8}\;s$, $\sin{\left(2\lambda_4\right)}=0$ and $\sin{\left(2\lambda_{\mu}\right)}=0$ to the observed energy distributions of excess events obtained in both operating modes.

Next, we investigate the event distribution as a function of polar angle.  We introduce the polar angle $\theta_{det}$ and the azimuthal angle $\phi_{det}$ between the muon neutrino beam axis and the detected photon's 3-momentum. This definition is related to the previously introduced polar angle $\theta_{\gamma}$ for a given polar angle $\theta$:
\begin{equation}\label{anglecos}
\cos{\theta_{\gamma}}=\frac{\cos{\theta_{det}}\cos{\theta}+\sin{\theta_{det}}\sin{\theta}\cos{\phi_{det}}-v}{1-v\left(\cos{\theta_{det}}\cos{\theta}+\sin{\theta_{det}}\sin{\theta}\cos{\phi_{det}}\right)}\;,
\end{equation}
To define the predicted number of events, we need to cast Eqs.\;\eqref{decayproba} and \eqref{decayprobb} in terms of new angles $\left(\theta_{det},\phi_{det}\right)$.
The integration measure is transformed as
\begin{equation}\label{anglemeasure}
d\!\cos{\theta_{\gamma}}d\phi_{\gamma}=\frac{\left(1-v^2\right)d\!\cos{\theta_{det}}d\phi_{det}}{\left(1-v\left(\cos{\theta_{det}}\cos{\theta}+\sin{\theta_{det}}\sin{\theta}\cos{\phi_{det}}\right)\right)^2}\;.
\end{equation}
Using Eqs.\;\eqref{anglecos} and \eqref{anglemeasure}, we transform the decay rates \eqref{decayproba} and \eqref{decayprobb} into new variables $\theta_{det}$ and $\phi_{det}$. Analogously to the previous case, we introduce two functions:
\begin{equation}\label{polardistu}
\begin{split}
\mathcal{U}^{mode,particle}_{[\cos{\theta_{min}},\cos{\theta_{max}}]}=\left(\kappa_{\mu 4}^{tr}\right)^2\frac{\alpha}{8\pi}\mathcal{N}^{mode}_{p.o.t.}N_{C}\int \frac{\left(1-v^2\right)dE\,d\!\cos{\theta}\,d\!\cos{\theta_{det}}d\phi_{\det}}{\left(1-v\left(\cos{\theta_{det}}\cos{\theta}+\sin{\theta_{det}}\sin{\theta}\cos{\phi_{det}}\right)\right)^2}\times \\ \times E^3\sqrt{E^2-m_s^2}\epsilon_{\gamma}\rho_{mode}^{particle}\!\left(E\right) \frac{G^2\left(t\right)}{t^2}\left(\left(1+ v\right)^3\left(1-\cos\theta\right)+\left(1-v\right)^3\left(1+\cos\theta\right)\right)P\left(x\right) \times \\
\times \Theta\!\left(1+v\left(\cos{\theta_{det}}\cos{\theta}+\sin{\theta_{det}}\sin{\theta}\cos{\phi_{det}}\right)-\frac{2\sqrt{1-v^2}E_{min}}{m_s}\right)\;,
\end{split}
\end{equation}
\begin{equation}\label{polardistv}
\begin{split}
\mathcal{V}^{mode,particle}_{[\cos{\theta_{min}},\cos{\theta_{max}}]}=\left(\kappa_{\mu 4}^{tr}\right)^2\frac{\alpha}{8\pi}\mathcal{N}^{mode}_{p.o.t.}N_{C}\int \frac{\left(1-v^2\right)dE\,d\!\cos{\theta}\,d\!\cos{\theta_{det}}d\phi_{\det}}{\left(1-v\left(\cos{\theta_{det}}\cos{\theta}+\sin{\theta_{det}}\sin{\theta}\cos{\phi_{det}}\right)\right)^3}\times \\ \times \left(\cos{\theta_{det}}\cos{\theta}+\sin{\theta_{det}}\sin{\theta}\cos{\phi_{det}}-v\right)E^3\sqrt{E^2-m_s^2}\epsilon_{\gamma}\rho_{mode}^{particle}\!\left(E\right) \frac{G^2\left(t\right)}{t^2}\times \\ \times \left(\left(1+ v\right)^3\left(1-\cos\theta\right)-\left(1-v\right)^3\left(1+\cos\theta\right)\right)P\left(x\right) \times \\
\times \Theta\!\left(1+v\left(\cos{\theta_{det}}\cos{\theta}+\sin{\theta_{det}}\sin{\theta}\cos{\phi_{det}}\right)-\frac{2\sqrt{1-v^2}E_{min}}{m_s}\right)\;,
\end{split}
\end{equation}
where we integrate over $\phi$ from $0$ to $2\pi$  and over $\cos\theta_{det}$ from $\cos{\theta_{max}}$ to $\cos{\theta_{min}}$, $\Theta$ is the step function, which selects events with observed photon energies $E_{\gamma}>E_{min}=0.2\;\mathrm{GeV}$. Then, the theoretical prediction for the number of events vs. $\cos{\theta}_{det}$ with observed photon energies $E_{\gamma}>E_{min}$ is
\begin{equation}\label{numberofeventspolar}
\begin{split}
\mathcal{N}^{mode}_{[\cos{\theta_{min}},\cos{\theta_{max}}]}&=\left(1-\sin{\left(2\lambda_{\mu}\right)}\right)\left(\mathcal{U}^{mode,particle}_{[\cos{\theta_{min}},\cos{\theta_{max}}]}-\sin{\left(2\lambda_{4}\right)}\mathcal{V}^{mode,particle}_{[\cos{\theta_{min}},\cos{\theta_{max}}]}\right)+\\
&+\left(1+\sin{\left(2\lambda_{\mu}\right)}\right)\left(\mathcal{U}^{mode,particle}_{[\cos{\theta_{min}},\cos{\theta_{max}}]}+\sin{\left(2\lambda_{4}\right)}\mathcal{V}^{mode,particle}_{[\cos{\theta_{min}},\cos{\theta_{max}}]}\right)\;.
\end{split}
\end{equation}

Figure \ref{figangulardist} demonstrates that almost all photons would be produced in the forward direction. To properly constrain $\kappa_{\mu 4}$ for polar angle distributions, one needs to know the uncertainties for the excess events. However, we did not find these estimates in the literature and thus do not use the polar angle distribution to place limits on $\kappa_{\mu 4}$. Note in passing that the excess distribution of $\cos\theta_{det}$ remains intact with a decrease in sterile neutrino lifetime (e.g., if other decay modes are introduced). Indeed, a decreased lifetime would simply add more energetic photons and increase the peak in the forward direction because function \eqref{averdecay} suppresses high energy particles more strongly than low energy particles and has a limit equal to one for large arguments.

\section{Analysis methods and results}\label{sectionanalysisandresults}
By fitting theoretically predicted numbers of events \eqref{numberofeventspolar} to the excess distributions, we see that the anomalies \cite{miniboone0812excess,miniboone1007excess,miniboone1207excess} cannot be entirely explained only by the effective electromagnetic dipole interaction \eqref{inter1}. For this reason, we adopted the following methods: (\!{\it i}) work on constraints and not on the expected values of parameters, which will explain the anomalies and (\!{\it ii}) analyze each bin and each mode independently. 

We calculated the numbers of events as functions of $\kappa^{tr}_{\mu 4}$, $\sin\lambda_{\mu}$ and $\sin\lambda_{4}$ for all bins and modes using Eq. \eqref{numberofevents}. After these calculations, we investigated the predicted numbers of events as functions of $\sin\lambda_{\mu}$, $\sin\lambda_{4}$ and $m_s$ using the known uncertainties of the experimental results for each bin and mode. In each bin, we constrained $\kappa^{tr}_{\mu 4}\left(\sin\lambda_{\mu}, \sin\lambda_4, m_s\right)$ at the 95\% C.L.. Finally, we constrained $\kappa^{tr}_{\mu 4}$ for a given $m_s$ by the minimum value over all bins and modes and the maximum value over the continuous variables $\sin\lambda_{\mu}$ and $\sin\lambda_{4}$ of the constraints on $\kappa^{tr}_{\mu 4}\left(\sin\lambda_{\mu}, \sin\lambda_4, m_s\right)$.

\begin{figure}[!h]
\begin{tabular}{cc}
 \includegraphics{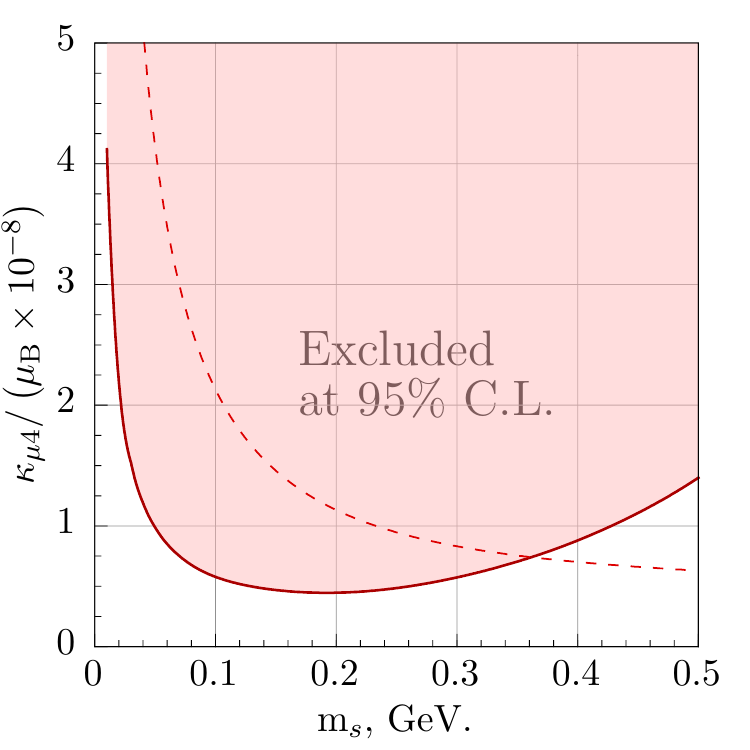} & \includegraphics{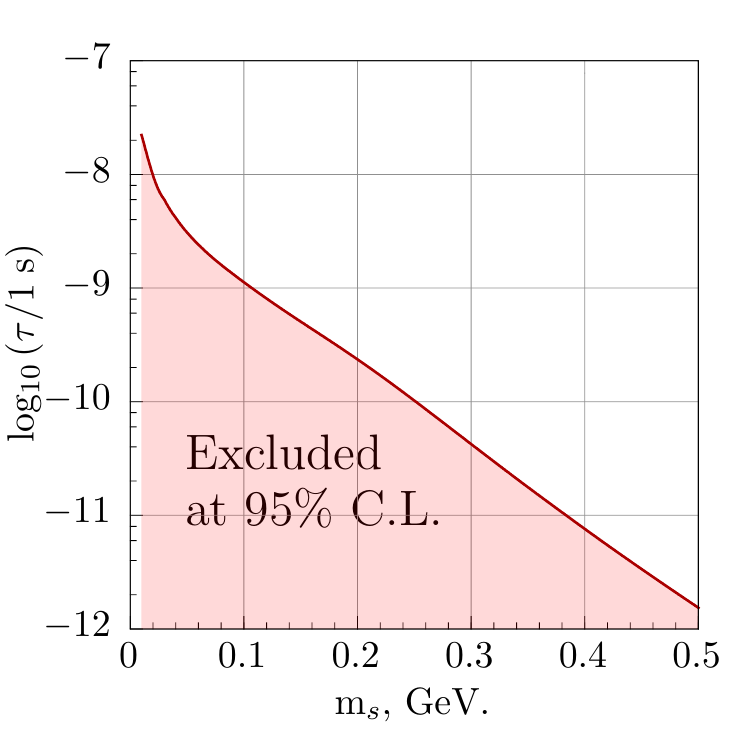}
\end{tabular}
\caption{{\small In the left plot, the shaded area in $\left(m_s,\; \kappa^{tr}_{\mu 4}\right)$ is excluded at the 95\% C.L. from our analysis. The region above the dashed line was excluded by Ref.\;\cite{gninenkokrasnikov1998}. In the right plot,  we present constraints on the sterile neutrino lifetime obtained from constraints on $\kappa_{\mu4}^{tr}$.}}
\label{consfig}
\end{figure}

The upper limit on $\kappa^{tr}_{\mu 4}$ at the 95\% C.L. is shown in Fig.\;\ref{consfig} (left), and  we present constraints on the sterile neutrino mean life time in Fig.\;\ref{consfig} (right), which are derived from the upper limit on $\kappa^{tr}_{\mu 4}$ (provided  Eq.\;\eqref{approxrel}). We see that data from the MiniBooNE detector allows us to improve constraints \cite{gninenkokrasnikov1998} for the mass range $m_s<350\; \mathrm{MeV}$.

Now let us discuss a modification of the previous analysis to models with more sterile neutrinos ($N_{\nu}\geq 5$). In particular, $\nu_s$ may decay into $\gamma$ and new sterile neutrino(s) $\nu_{s,j}$ ($j\geq5$) if it is kinematically allowed, i.e., $m_s>m_{s,j}$. Eqs.\;\eqref{decayproba} and \eqref{decayprobb} take different forms:
\begin{equation}\label{decayprob2a}
\begin{split}
\frac{d\Gamma^{\nu}_{\pm}}{d\phi_{\gamma}d\cos\theta_{\gamma}}&=\frac{1}{32\pi^2}m_s^3\sum_{i=1}^3\left(\kappa^{tr}_{4i}\right)^2\left(1\mp\sin\left(2 \lambda^{tr}_{4i}\right)\cos{\left( \phi^{tr}_{4i}- \chi^{tr}_{4s,j}\right)}\cos{\left(\theta_\gamma\right)}\right)  +\\
&+\frac{1}{32\pi^2}m_s^3\sum_{j=5}^{N_{\nu}}\Theta\left(m_s-m_{s,j}\right)\left(1-\frac{m_{s,j}^2}{m_s^2}\right)^3\left(\kappa^{tr}_{4s,j}\right)^2\times \\
&\times \left(1\mp\sin\left(2 \lambda^{tr}_{4s,j}\right) \cos{\left( \phi^{tr}_{4s,j}- \chi^{tr}_{4s,j}\right)}\cos{\left(\theta_\gamma\right)}\right)\;,
\end{split}
\end{equation}
\begin{equation}\label{decayprob2b}
\begin{split}
\frac{d\Gamma^{\overline{\nu}}_{\pm}}{d\phi_{\gamma}d\cos\theta_{\gamma}}&=\frac{1}{32\pi^2}m_s^3\sum_{i=1}^3\left(\kappa^{tr}_{4i}\right)^2\left(1\pm\sin\left(2 \lambda^{tr}_{4i}\right)\cos{\left( \phi^{tr}_{4i}- \chi^{tr}_{4s,j}\right)}\cos{\left(\theta_\gamma\right)}\right)  + \\
&+\frac{1}{32\pi^2}m_s^3\sum_{j=5}^{N_{\nu}}\Theta(m_s-m_{s,j})\left(1-\frac{m_{s,j}^2}{m_s^2}\right)^3\left(\kappa^{tr}_{4s,j}\right)^2\times \\
&\times\left(1\pm\sin\left(2 \lambda^{tr}_{4s,j}\right)\cos{\left( \phi^{tr}_{4s,j}- \chi^{tr}_{4s,j}\right)}\cos{\left(\theta_\gamma\right)}\right)\;.
\end{split}
\end{equation}
The energy of the outgoing photon as a function of $E$ and $\cos\theta$ is determined through
\begin{equation}\label{gammaenfor2}
E_{\gamma}=\frac{m_s}{2}\left(1-\frac{m_{s,j}^2}{m_s^2}\right)\frac{1+v\cos\theta}{\sqrt{1-v^2}}\;.
\end{equation}
Consequently, we have to introduce modifications of the cutoff function  $\theta_{cut}$ (see Eq.\;\eqref{cutangle1}), which now depends on $m_{s,j}$:
\begin{equation}
\cos\theta_{cut}^{s,j}\!\left(E\right)=\cos\theta_{cut}\!\left(\frac{E}{\left(1-\frac{m_{s,j}^2}{m_s^2}\right)}\right)\;.
\end{equation}

We are ready now to discuss the second scenario mentioned in the Introduction \cite{masip1210}, which suggested an explanation for the MiniBooNE and LSND anomalies. Initially, the scenario allowed flavor mixing between sterile and muon neutrinos, but such mixing was found to be inconsistent with kaon decays \cite{duk2012}. Then, the authors of  \cite{masip1210} introduced two sterile Majorana neutrinos with $m_s=50\;\mathrm{MeV}$ and $5\;\mathrm{MeV}< m_{s,5}<10\;\mathrm{MeV}$ and two transition magnetic dipole moments $\mu^{tr}_{\mu 4}=2.4\times 10^{-9}\mu_B$ and $\mu^{tr}_{45}=2.4\times 10^{-8}\mu_B$; the latter moment saturates the heavy sterile neutrino decay rate via $\nu_s\rightarrow\nu_{s,j}\gamma$. Equation\;\eqref{gammaenfor2} implies that the massless approximation is reasonable for such a hierarchy of $m_s$ and $m_{s,5}$. To test the case with two sterile neutrinos, we fix the proposed relation between magnetic moments as $\mu^{tr}_{\mu 4}=0.1\times\mu^{tr}_{45}$. Using Eq.\;\eqref{numberofevents}, we place the constraint $\mu^{tr}_{45}<2.1\times10^{-8}\mu_B$ (95\% C.L.) for Majorana neutrinos, which is marginally consistent with the given value of $\mu^{tr}_{45}$. However, such a model is evidently disfavored because the additional source of photons would give a peak in the forward direction. Extensions of the obtained constraints to models with several neutrinos are straightforward. More careful analysis is required to test models with almost degenerate (in mass) sterile neutrinos. In particular, in the model with two degenerate neutrinos $m_4 \approx m_{s,5}$ for $\left(\kappa_{45}^{tr}\right)^2\left(1-m_{s,5}^2/m_s^2\right)^3 \gg \left(\kappa_{\mu 4}^{tr}\right)^2$ most photons from the decay $\nu_4\rightarrow \nu_{s,5}\gamma$ could have energies comparable or below the  lower energy cut off at 200 MeV. In such a specific case, a sterile neutrino could avoid our constraints for $\kappa_{45}^{tr}$ and $\kappa_{\mu 4}^{tr}$.However, such a situation is excluded as an explanation of the excess because experimental data shows a smooth angular dependence \cite{miniboone1207excess}. 

\section{Conclusion}\label{sectionconclusion}

Combined analyses from different focusing regimes of the MiniBooNE detector have allowed us to investigate models with heavy sterile neutrinos. In particular, we renewed and generalized the constraints on electromagnetic transition dipole moments. The main results are presented in Fig.\;\ref{consfig}. The obtained analytical formulas are rather general and are applicable to other neutrino experiments.

We have shown that sterile neutrinos of mass $10\;\mathrm{MeV}<m_s<500\; \mathrm{MeV}$ that have transition dipole moments and do not significantly mix with active neutrinos of mass in the region $10\;\mathrm{MeV}<m_s<500\;\mathrm{MeV}$ cannot explain anomalies if one regards the MiniBooNE polar angle distribution of the anomalous events. 
\section*{Acknowledgments}

We thank D.~S.~Gorbunov and S.~N.~Gninenko for stimulating questions and fruitful discussions. The work was partially supported by the grant from the Ministry of Education and Science No. 8412, the grant from the President of the Russian Federation NS-5590.2012.2 and the RFBR grant 13-02-01127.

\bibliographystyle{apsrev}
\bibliography{radionov_constraints2013}	

\end{document}